\documentclass[preprint,12pt,authoryear]{elsarticle}

\usepackage{amssymb}
\usepackage{amsmath}

\usepackage{graphicx}
\usepackage{subcaption}
\usepackage{txfonts}
\usepackage{lscape}
\usepackage{booktabs}
\usepackage{multirow}
\usepackage{multicol}
\usepackage{lineno}
\usepackage{bm}
\usepackage{color}
\usepackage{orcidlink}
\newcommand{\bmath}[1]{\bm{#1}}

\newcommand\Dima[1]{#1}

\newcommand{\aap}{A\&A}
\newcommand{\icarus}{Icarus}
\newcommand{\mnras}{MNRAS}

\journal{Icarus}

\begin{document}

\begin{frontmatter}



\title{A practical re-weighting scheme of data fitting:
application to asteroids orbit determination with \textit{Gaia}}





\author[labelOP,labelIAA,labelUW]{Dmitri.~E.~Vavilov \, \orcidlink{0009-0007-1972-5975}}
\author[labelOP]{Ziyu.~Liu \, \orcidlink{0009-0004-6815-5794}}
\author[labelOP]{Daniel.~Hestroffer \, \orcidlink{0000-0003-0472-9459}}
\author[labelIPSA,labelOP]{Josselin.~Desmars \, \orcidlink{0000-0002-2193-8204}}

\affiliation[labelOP]{organization={LTE, Observatoire de Paris, Université PSL,
Sorbonne Université, Université de Lille, LNE, CNRS},
            addressline={61 Avenue de l’Observatoire},
            city={Paris},
            postcode={75014},
            country={France}}

\affiliation[labelIAA]{organization={Institute of Applied Astronomy,
Russian Academy of Sciences},
            addressline={Kutuzova emb. 10},
            city={St.\ Petersburg},
            country={Russia}}

\affiliation[labelUW]{organization={DiRAC Institute and the Department of Astronomy,
University of Washington},
            addressline={3910 15th Ave NE},
            city={Seattle},
            postcode={98195},
            state={WA},
            country={USA}}

\affiliation[labelIPSA]{organization={Institut Polytechnique des Sciences Avancées (IPSA)},
            addressline={63b Bd. de Brandebourg},
            city={Ivry-sur-Seine},
            postcode={94200},
            country={France}}

\begin{abstract}

The method of weighted least squares is widely used in parameter estimation problems such as asteroid orbit determination. A frequent difficulty is the realistic treatment of observational uncertainties, especially when combining heterogeneous datasets with very different precision. We propose a quick and simple reweighting scheme that adjusts the contribution of each measurement group to ensure a statistically consistent least-squares solution. The method consists of three steps: (i) estimating the error standard deviations for each observational subset, (ii) rescaling their weights by the corresponding variances, and (iii) performing a weighted least-squares fit using the adjusted weights. 
We apply this approach to heliocentric orbit fitting of asteroids using combined ground-based astrometry and high-precision \textit{Gaia} measurements. We validated the method by fitting each orbit to a restricted observation set and comparing its predictions with the complete set of measurements. For 7 objects, the reweighted solutions provide significantly improved agreement with older data. The most dramatic case is asteroid (21) Lutetia, where increasing the effective uncertainty of \textit{Gaia} observations by a factor of 17 yields a substantially better fit, indicating the importance of accounting for possible systematic biases in high-precision datasets.
We further apply the scheme to the recently discovered near-Earth asteroid 2024~YR$_4$, where we grouped observations by the visual magnitude. The reweighted orbit produces smaller uncertainty regions and a more stable solution, reducing predicted impact probabilities by roughly an order of magnitude. All computed probabilities remain below 0.5\%, under the 1\% International Asteroid Warning Network (IAWN) alert threshold.
This reweighting procedure provides a practical way to combine measurements of heterogeneous quality, improving the reliability of least-squares solutions in orbit determination and impact-risk assessment. The method is general and can be readily applied to other parameter estimation problems involving mixed-precision data.

\end{abstract}





\begin{highlights}
\item We propose a simple reweighting scheme that divides observations into groups, estimates the rms for each group, and adjusts the weights according to their measured precision.
\item This reweighting method produces more precise asteroid orbits and improves the quality of future-position predictions.
\item Applying the scheme to asteroid 2024~YR$_4$ resulted in smaller orbital uncertainties and more reliable impact-probability estimates, keeping all values safely below the 1\% IAWN alert threshold.
\end{highlights}


\end{frontmatter}



\section{Introduction}

The least-squares method \citep{1809tmcc.book.....G} is widely used in all types of physical sciences to estimate the parameters of the system from real observations. It is based on minimizing the sum of the squared differences between the observed and predicted values, which corresponds to maximizing the likelihood of the parameters when observational errors follow a (multivariate, independent, identically distributed) Gaussian distribution. One of the key components of the least-squares method is the proper assignment of weights to different types of observations, where the weights are the inverse of the dispersion of the corresponding errors. These weights directly affect the results of a fit, determining how much influence each observation has on the final solution. Inaccurate or improper weighting can lead to biased or suboptimal results, especially when combining observational data of different types with varying levels of precision. Besides, the variance-covariance matrix, which determines the uncertainty region of the fitted parameters, can show a significant variability with various weighting schemes. These bias in estimator and error in confidence level can strongly affect the collisional prediction and impact probability of Near-Earth asteroids, or the uncertainty of the ephemeris position, also important for asteroids' precovery or follow-up \citep{2024A&A...689A..49V}.


Previous studies have explored the importance of assigning appropriate weights to astrometric data for asteroid and comet orbit determination. For instance, \citet{2003Icar..166..248C} emphasized the critical role of weighting astrometric observations in orbital determination and proposed initial frameworks to adjust observation errors systematically. More recently, \citet{2015Icar..245...94F} and \citet{2017Icar..296..139V} demonstrated how proper weighting of astrometric data led to improvements in the predictability of asteroid orbits, reducing uncertainties in ephemeris predictions.

However, while much progress has been made, the optimal strategy for weighting astrometric data remains a problem needing more investigation. The differences in quality between classical astrometry, which is prone to larger errors, and modern space-based data, such as Gaia observation {\citep{tanga23}}, require a careful treatment that can reconcile these data sources without introducing additional biases. Incorrect weighting of observations can lead to the overestimation or underestimation of uncertainties, ultimately affecting the precision of orbital parameter solutions.


In this paper, we introduce a method for adjusting weights to data from different types and quality, to enhance the accuracy of the determined parameters, and apply it to astrometric observations of asteroids. Our method divides observations into distinct groups, determining the weight ratios between the groups while preserving the internal weight relation within each group. In Section~2 we describe our approach and in Section~3, we then present the validation on the case of combining ground-based and Gaia positional observations of asteroids for orbit determination. In Section~4 we apply our scheme on near-Earth asteroid 2024~YR$_4$ to obtain more precise orbits and impact probability estimates. The conclusion follows this section.

\section{The method}
\subsection{Weighted least square fit}

The least-squares method, initially developed by Carl Friedrich \citet{1809tmcc.book.....G}, is widely used for fitting a model to data by minimizing the sum of the squared differences between the observed values and the model predictions. In its weighted form, the method is particularly useful when the observations are of varying precision or variance (heteroscedasticity). In such cases, each observation is assigned a weight that reflects its uncertainty, ensuring that more reliable data points have a greater influence on the fit than less reliable ones.

In a weighted least-squares fit, we consider a set of observations $y_i$ (for $i = 1, 2, \dots, n$) corresponding to independent variables $t_i$ typically time and a model function $f(t, \bmath{X})$ that depends on a set of parameters $\bmath{X}$. We assume that the errors of observations $y_i$ are distributed by a Gaussian law with standard deviation $\sigma_i$, different for each observation. The parameters $\bmath{X}$, which maximize the likelihood, are found from minimizing the weighted sum of squared residuals, defined as:
\begin{equation}
S(\bmath{X}) = \sum_{i=1}^{n} w_i [y_i - f(t_i, \bmath{X})]^2,
\end{equation}
\noindent where $w_i$ represents the weight associated with the $i$-th observation ($w_i = 1 / \sigma_i^2$). 

To find the best-fit parameters $\bmath{X}$, we minimize the function $S(\bmath{X})$. This is done by taking the partial derivatives of $S(\bmath{X})$ with respect to each parameter $x_j$ and setting them equal to zero:
\begin{equation}
\frac{\partial S(\bmath{X})}{\partial x_j} = -2 \sum_{i=1}^{n} w_i [y_i - f(t_i, \bmath{X})] \frac{\partial f(t_i, \bmath{X})}{\partial x_j} = 0.
\end{equation}
Solving this system of equations (known as the conditional equations) provides the values of the parameters that minimize the weighted sum of squared residuals.

In practice, the derivatives $\frac{\partial f(t_i, \bmath{X})}{\partial x_j}$ are unknown, and the system is solved numerically by iterations. 


\subsection{Uncertainty adjustment parameter}
\label{sec:adjustment_param}

Let's assume that we have two groups of observations (with $N_1$ and $N_{2}$ number of observations, respectively) coming from two different sources. 
Determining the appropriate relative weights between these groups poses a challenge, in particular if the observations are of different nature (e.g. photometric and astrometric). In contrast, the relative weights within each individual group can be more stable and easier to estimate, as they are subject to a more consistent set of conditions and observational biases. This consistency within groups allows for a more reliable estimation of internal weights, making intra-group weighting a feasible task even when inter-group weighting remains complex.

We propose the following empirical approach of reweighting the observations to improve the accuracy of the resulted parameters. Let's assume that we already have the set of estimates of the weights of observations $\{w_i, i=1,...,n\}$.

\begin{enumerate}
    \item Decrease the weights of the \textbf{second} group of observations by a large factor, for instance, $10^8$ times (\Dima{$10^4$} times increase of the standard deviation) and fit the parameters yielding $\bmath{X}_1$.
    \item Decrease the weights of the \textbf{first} group of observations by $10^8$ times (while using the original weights of the second group) and fit the parameters yielding $\bmath{X}_2$.
    \item Compute K-factors defined as:

    \begin{equation}
        \begin{aligned}
            K_1 = & \sqrt{ \frac{\sum_{i=1}^{N_1}{w_i(O_i-C_i)_{1}^2}}{N_1 - m} } \\
            K_2 = & \sqrt{ \frac{\sum_{i=1}^{N_2}{w_i(O_i-C_i)_{2}^2}}{N_{2} - m} },
        \end{aligned}
        \label{eq:k_factors}
    \end{equation}
\noindent where vector $(O-C)_l$ is the ''observed minus computed'' for the l-th group of observations with ''computed'' made from $\bmath{X}_l$ parameters, and where $m$ is the size of vector $\bmath{X}$ (number of independent parameters).

In fact, $K^2 / w_i$ is an unbiased estimation of a standard deviation of $i$-th observation.

\item Adjust the weights of each group by dividing the weights of the first group by $K_1^2$ and the second group by $K_2^2$.

\item Fit the parameters with new weights obtained from the previous stage.
    
\end{enumerate}

If the original weights are adequate then $K_1$ and $K_2$ must be close to 1. Basically we estimate the uncertainty by the standard deviation of the observations in each of the group separately and adjust the weights according to it. One should note that this approach can be applied only if the number of observations in each group is large enough to have trustworthy estimates of standard deviations.

We want to stress that we do not set zero weights to one of the group and fit the parameters to obtain $K_1$ and $K_2$ on purpose. Firstly, it can help with the stability and convergence of iterations in the fitting. But most importantly, the proposed scheme can be used even when we combine observations, that are not enough to fit the parameters solely. For instance, when fitting orbits of binary asteroids and using a combination of photometric and astrometric observations. \Dima{Only} photometric observations \Dima{(visual magnitude vs time)} are not enough to find the orbital parameters, however, with the proposed scheme, it is possible to estimate a proper relative weight of photometric observations with respect to positional ones. The scaling factor ($10^8$ here), while arbitrary, must be large enough so that the observational errors and possible systematic deviations will be  smaller than standard deviations after the rescaling. For example, using $10^8$ means we enlarge the assumed observational error by a factor of 10,000, which should be enough in most of the applicaiton cases studied below.

In practice, we can simplify the proposed scheme: If one has some initial estimations of the weights that  should not deviate from the `real' values by an order of magnitude, one can merge step \#1 and step \#2. Hence, we fit the parameters $\bmath{X}$ with the primordial weights and use it to then compute $K_1$ and $K_2$. This process is faster and allows us to combine more groups of observations without additional computational cost.

\section{Application on using ground-based and Gaia observations}

\begin{table}[h]

\scriptsize
\caption{Information of the 15 test asteroids.}
\begin{tabular}{c|cccc|ccccc}

\hline  \hline 
\multirow{2}{*}{\begin{tabular}[c]{@{}c@{}}Asteroid\\ Number\end{tabular}} & \multicolumn{4}{c|}{Full Set}              & \multicolumn{5}{c}{Subset Fitted}                                 \\ \cline{2-10} 
                                                                           & Arc of Observation       & ${N_{GB}}$ & ${N_{Gaia}}$ & ${\delta\chi^2}$ & Arc of Observation       & ${N_{GB}}$ & ${N_{Gaia}}$ &$ {k_{GB}}$    & ${k_{Gaia}}$     \\ \hline \hline 
1917                                                                       & 1954-05-06 -- 2024-12-12 & 2936  & 380 &  1.34     & 2016-01-13 -- 2016-03-31 & 121   & 58      & 0.62 & 0.74 \\
3199                                                                       & 1982-09-13 -- 2023-08-08 & 2387  & 319 &  1.35     & 2018-10-24 -- 2019-02-25 & 110   & 45      & 1.71 & 0.82   \\
1036                                                                       & 1924-10-23 -- 2024-07-06 & 8558  & 433 &  124.84    & 2016-12-10 -- 2017-04-29 & 397   & 117     & 0.26 & 1.13 \\
7358                                                                       & 1959-12-02 -- 2024-07-07 & 2229  & 265 &  1.14    & 2018-09-11 -- 2019-03-07 & 190   & 33      & 1.24 & 0.69  \\
3122                                                                       & 1979-03-09 -- 2024-07-06 & 5730  & 245 &  1.00    & 2019-03-01 -- 2019-10-06 & 173   & 62      & 1.07 & 0.92  \\
3288                                                                       & 1982-02-28 -- 2023-12-07 & 2238  & 227 &  3.35    & 2018-09-08 -- 2019-01-05 & 129   & 53      & 1.42 & 0.72  \\
21                                                                         & 1866-04-06 -- 2024-04-16 & 5877  & 493 &  5.97     & 2018-05-05 -- 2018-09-21 & 157   & 239     & 0.75 & 17.16  \\
1916                                                                       & 1953-09-01 -- 2024-05-03 & 1972  & 366 &  4.41    & 2017-12-08 -- 2018-12-31 & 246   & 84      & 0.68 & 0.81  \\
887                                                                        & 1918-01-03 -- 2024-09-30 & 2926  & 411 &  1.14    & 2019-03-01 -- 2019-08-27 & 62   & 90      & 1.06 & 0.71  \\
3102                                                                       & 1981-08-21 -- 2024-03-05 & 2336  & 169 &  1.04    & 2019-08-04 -- 2020-02-15 & 433   & 44      & 1.12 & 0.78  \\
5751                                                                       & 1989-03-10 -- 2024-09-29 & 1999  & 267 &  294.88    & 2015-11-01 -- 2016-02-21 & 80    & 53      & 0.34 & 0.98 \\
6086                                                                       & 1960-11-17 -- 2024-04-21 & 4213  & 300 &  1.13    & 2019-09-07 -- 2019-12-30 & 145   & 48      & 0.72 & 0.80  \\
7638                                                                       & 1969-01-15 -- 2024-09-08 & 3417  & 608 &  1.00    & 2019-05-06 -- 2019-09-30 & 80   & 101      & 1.02 & 0.74  \\
1862                                                                       & 1930-12-13 -- 2024-01-31 & 3083  & 243 &  0.82    & 2016-10-26 -- 2017-01-30 & 91   & 88      & 0.72 & 0.77  \\
7345                                                                       & 1969-10-15 -- 2024-05-02 & 2177  & 429 &  0.69    & 2018-01-25 -- 2018-08-02 & 90   & 192      & 0.87 & 0.77  \\\hline


\end{tabular}
 $N_{GB}$ and $N_{Gaia}$ stand for the number of ground-based and Gaia observations correspondingly.
 $K_{GB}$ and $K_{Gaia}$ are the $K$ factors computed using Eq.~(\ref{eq:kgaia}) to adjust the weight of the observations.
  $\delta \chi^2$ is the ratio of the weighted chi-squared values between the two orbits after and before applying the new weighting. $\delta \chi^2 > 1$ means the new orbit better represents the full observation set.

\label{tb:res}
\end{table}

\subsection{Heliocentric orbit fitting}


To validate the proposed method, we applied it to the heliocentric orbit fitting of asteroids. In such cases, weighted least-squares fitting is a widely used technique, but due to the large diversity of observation sources, techniques, and quality, the appropriate determination of uncertainties (and consequently the weights) can be challenging. To address this, scaling factors $K_1$ and $K_2$ are introduced to adjust the weights, improving the fit quality.

The heliocentric orbit fitting in our study was performed using the Numerical Integration of the Motion of Asteroids (NIMA) program, as described in \citet{NIMA}. This program propagates asteroid orbits and calculates partial derivatives while accounting for perturbations from major planets, the Moon, four additional massive asteroids (Ceres, Vesta, Pallas, Hygiea), and relativistic corrections. NIMA employs the weighted linear least-squares method described above to compute differential corrections to the initial state vector, optimizing the fit to the observational data.

To perform the orbit adjustment, NIMA first extracts the observation file for the target object from the Minor Planet Centre (MPC), and assigns a fixed uncertainty according to the observatory and reference catalog for each observation by referring to \citet{2015Icar..245...94F}. Next, the \textit{Gaia} observations of the Focused Product Release \citep{david23} are extracted directly from the Gaia archive along with its uncertainty. The uncertainty value follows the \textit{Gaia} error model, which provides systematic and random error estimates as outlined in Chapter~8 of the \textit{Gaia} Data Release~3 documentation \citep{gdr3-22}.

These uncertainties are thus taken into account in the orbit calculation by adding the weighting scheme. 
\begin{equation}
    \mathbf{W} = \frac{1}{1-\rho^2}
    \begin{bmatrix}
    \frac{1}{{\sigma^2_{\alpha\cos{\delta}}}} & \frac{-\rho}{\sigma_{\alpha\cos{\delta}}\sigma_{\delta}}\\
    \frac{-\rho}{\sigma_{\alpha\cos{\delta}}\sigma_{\delta}} & \frac{1}{{\sigma^2_{\delta}}}
    \end{bmatrix}
\end{equation} 

\noindent where $\rho$ is the correlation between the right ascension RA and declination Dec (non-zero for Gaia observations), and $\sigma_{\alpha\cos{\delta}}$, $\sigma_{\delta}$ are the corresponding uncertainty, respectively. 

And the 
final solution is given by:
\begin{equation}
    \bmath{X} = (\bmath{B}^T\mathbf{W}\bmath{B})^{-1} \ \bmath{B}^T \mathbf{W}\bmath{Y}
\end{equation} 

\noindent where $\bmath{Y=(O-C)}$ represents the difference between observed and computed positions, $\bmath{B}$ is the partial derivative function with respect to the state vector, $\bmath{X}$ is the correction to apply to the state vector for the improved orbit, and $T$ denotes the matrix transpose operation. Additionally, the uncertainty and confidence ellipsoid of the solution is given by the variance-covariance matrix $(\bmath{B}^T\mathbf{W}\bmath{B})^{-1}$.

\subsection{Data processing}\label{sec:data}

In order to show the improvement of the proposed scheme for weights adjustment, we select a small observation arc (typically several months), containing both Gaia and ground-based observations, that will be used for the orbit computation. We fit the orbit using the default weights, and also fit the orbit with the improved weights from the scheme described in Section~\ref{sec:adjustment_param}. Then we propagate these two orbits through the whole set of observations. Obviously, one expects that both orbits describe well the subset of observations which were used in the orbital fit. However, one expects to see some discrepancy, especially for the oldest observations or observations far from the selected observational arc.

The $K$ factors for ground-based and Gaia observations are computed respectively by: 
\begin{equation}
    \begin{aligned}
        K_{GB} = & \sqrt{\frac{1}{2N_{GB}-6} \sum_i{\frac{(O_i-C_i)^2_{\alpha\cos{\delta}}}{\sigma_{\alpha\cos{\delta},\ i}^2} + \frac{(O_i-C_i)_{\delta}^2}{\sigma_{\delta,\ i}^2}}} \\
        K_{Gaia} = & \sqrt{\frac{1}{2N_{Gaia}-6} \sum_i{\frac{(O_i-C_i)_{AL}^2}{\sigma_{AL,\ i}^2} + \frac{(O_i-C_i)_{AC}^2}{\sigma_{AC,\ i}^2}}}
    \label{eq:kgaia}
    \end{aligned}
\end{equation}

\noindent where $N_{GB}$ and $N_{Gaia}$ are the number of ground-based and Gaia observations, respectively (each containing two measured parameters, either RA Dec, or AL AC \footnote{\Dima{Gaia measures source positions in a scanning mode, where observations are recorded along the direction of the satellite’s rotation (along-scan, AL) and perpendicular to it (across-scan, AC). The measurement precision is much higher in the AL direction, since positions are determined from precise timing of the signal as the source crosses the CCD, while AC positions are derived with significantly lower spatial resolution.}}), correspondingly, $(O_i-C_i)_{\alpha\cos{\delta}}$ and $(O_i-C_i)_{\delta}$ are the 'observed-minus-computed' for the $i$-th observation for right ascension and declination.  $(O_i-C_i)_{AL}$ and $(O_i-C_i)_{AC}$ are the 'observed-minus-computed' for the $i$-th observation 'along scan' and 'across scan' (the two independent directions in Gaia observations; see section 2 in \cite{gaiadr2}). $\sigma_{\alpha\cos{\delta},\ i}$, $\sigma_{\delta,\ i}$, and $\sigma_{AL,\ i}$, $\sigma_{AC,\ i}$ are the assigned standard deviations of $i$-th observation in right ascension and declination in case of ground-based observations, or 'along scan' and 'across scan' in case of Gaia observations. 

The full procedure of the experiment -- as described in \ref{sec:adjustment_param} -- is to first use the original assigned weight to calculate the orbit, then decrease the weight for Gaia observations in the same subset (for example, by dividing the Gaia weight by \Dima{$10^8$}), then run the adjustment again to obtain the residuals, from which we calculate $K_{GB}$, and vice versa to calculate $K_{Gaia}$. We then apply these two factors to the initial weight to obtain a new orbit to compare with the first one. An alternative way of proceeding is to use the residuals of Gaia and GB observations from the first run directly to calculate the two $K$ factors. We show the results of using these two approaches in Table \ref{tb:1917} for one example asteroid. 
\begin{table}[]
    \centering
        \caption{Values obtained from different procedures presented in Section \ref{sec:data} for one example asteroid (1917) Cuyo}
    \begin{tabular}{|c|c|c|}
    \hline 
         1917 &Full procedure&Simplified Procedure  \\
         \hline 
         $K_{Gaia}$ &0.72&0.74\\
         $K_{GB}$ &0.58&0.62\\
         $\delta\chi^2$&1.34&1.34\\
          \hline 
    \end{tabular}
    \label{tb:1917}
\end{table}
One can see that the values we obtain from these approaches are very similar. Consequently, we will use the simplified procedure in subsequent runs.

\begin{figure*}[]
    \centering
    \includegraphics[width=0.93\textwidth,height=\textheight,keepaspectratio]{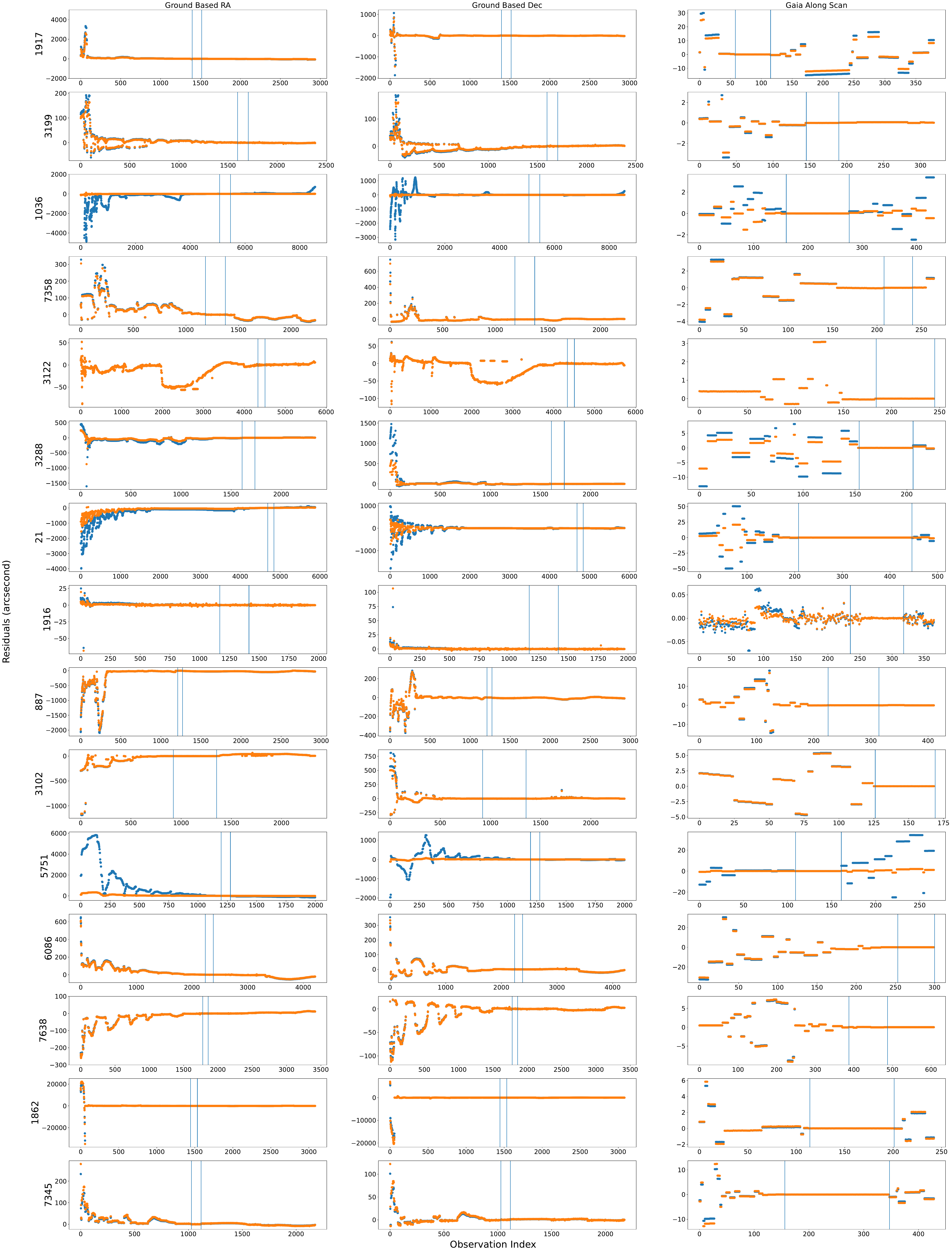}
    \caption{Comparison for the propagated O-C using the orbit before and after applying the new weighting scheme. The orange points show the results using the new weights and the blue points are for the original weights. The x-axis of the graph is the observed index for a better overview of the points. The vertical blue lines define the sub-set of observations selected for performing the orbit fitting (see text). First and second columns are O-C in right ascension and declination of ground-based observations; the third column is in Gaia observations along-scan.}
    \label{fig:comparison}
\end{figure*}

\subsection{Results}

We selected a sample of 15 asteroids to test the proposed reweighting scheme (see Table \ref{tb:res}). The selection process presented several challenges. On the one hand, each asteroid needed to have a substantial number of observations, including historical data, to adequately highlight differences between the computed orbits \Dima{(in this work we were looking for more than 2000 observations in total per object and at least 30 years of observational arc)}. On the other hand, it was crucial to carefully define the subset of observations used for orbit fitting. The observation window needed to be short enough (limited to several months) to ensure that the two resulting orbits would not both be overly precise, which would mask the differences between them. At the same time, the subset had to include a sufficient number of ground-based and \textit{Gaia} observations, typically on the order of several dozen (and $\ge 30$).
This meticulous selection process is time-consuming, which constrained our sample size to 15 asteroids, with different semi-major axis, eccentricity, sizes, etc. Nevertheless, we believe that this carefully curated sample is sufficiently large to demonstrate the effectiveness and importance of the proposed reweighting technique.


In Table~\ref{tb:res}, we present details of the selected asteroids, including the observation arcs, the number of observations, and the $K$-values computed using the simplified re-weighting scheme, detailed in the previous Section. Additionally, we provide $\delta \chi^2$, which represents the ratio of the weighted root-mean-square (rms) errors $\chi^2 = \sum w_i(O_i-C_i)^2$ between the two compared orbits, computed over the complete observational set. 
While computing this, we use the original weights, intentionally giving the original orbit a small advantage. A value of $\delta \chi^2 > 1$ indicates that the orbit computed with the new weights better represents the complete observation set.

Figure~\ref{fig:comparison} shows a visual comparison of how the two orbits fit the full set of observations. For seven asteroids (1917, 3199, 1036, 3288, 21, 1916, and 5751), the proposed reweighting scheme yields significant improvement that can be seen in Fig.~\ref{fig:comparison}. 
For the remaining asteroids, we observe slight improvements for four asteroids (7358, 887, 3102 and 6086), no noticeable difference for two of them (3122 and 7638), and a minor deterioration for another two (1862 and 7345). Overall, these results demonstrate that the proposed reweighting scheme either significantly improves the orbit fit, or leaves it mostly unchanged.

A particularly notable case is asteroid (21) Lutetia, where the improvement was substantial ($ \delta \chi^2 = 294 $). For this asteroid, the computed $K_{Gaia} = 17$ indicates that the Gaia observations were considerably less precise than initially assumed. As discussed by \citet{tanga23}, for large objects like Lutetia, Gaia data (in DR3 and FPR releases) can suffer from biases due to the (uncorrected) offset between the photocenter and the barycenter, in addition to other limitations for bright and large moving Solar System objects. The large $K_{Gaia}$ value is a good indicator that such phase effects should otherwise be considered in the orbit fitting process. The new solution hence appears to be more robust to such biases that are not included in the error model. However, we want to stress that even the reweighted solution turned out to be more stable it still cannot replace the proper handle of the systematic effects. 

Based on the demonstration, we show that when combining ground-based and modern Gaia data, 
applying the proposed $K$ weighting factors for different groups of observations provides more reliable results on orbits computation. Further, when dealing with the whole set of observations, it is recommended to divide it into subsets based on instrumentation and epochs, containing enough data (typically~$\ge 30$) for computing the respective weighting factor.


\section{Application to NEO 2024~YR$_4$}


%

While the proposed weighting scheme is reliable for combining ground-based with the high-precision space-based observations from Gaia, it remains general and applicable to other observations of a target made with different instruments and telescopes, and of different precision or accuracy. Besides, we can apply the proposed scheme to the orbit determination of the recently discovered near-Earth asteroid 2024~YR$_4$, and subsequently compute its impact probability (IP). This Apollo-type asteroid was discovered after its perihelion passage, on 27 December 2024 at the ATLAS station in Chile. It is not classified as a PHA because it is smaller in size than 140\,m. Nevertheless, it rapidly showed an IP larger than  1\%, raising an alert to the IAWN network (International Asteroid Warning Network) in late January 2025. Such alert is requiring additional data to secure the orbit and size, and possible threat in year 2032, just before its conjunction with the Sun. With 504 astrometric data taken over several months available at the Minor planet center, the risk has been now lowered to zero. 

\begin{figure}
    \centering
    \includegraphics[width=1\linewidth]{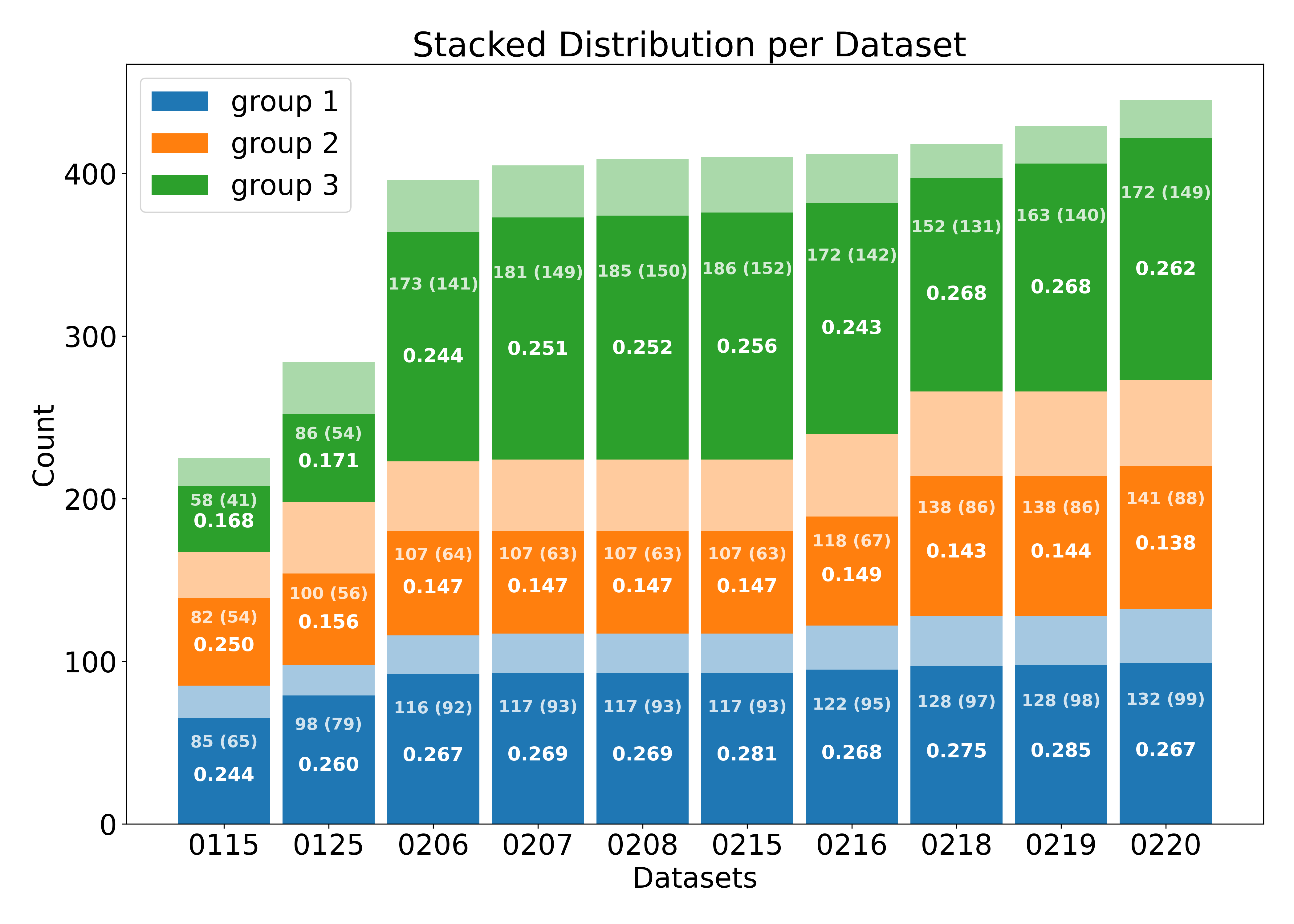}
    \caption{Distribution of the observations of 2024~YR$_4$ by 3 groups used for revising the weights of observations. The integer number indicates the total amount of observations in the group, the number in the brackets is the number of observations used after rejection and the real number shows the $K$-value for the orbit fits with outlier rejections by 3$\sigma$ rule. }
    \label{fig:groups_of_observations}
\end{figure}

Here we propose to apply our weighting scheme to the orbit fitting, and check for changes in the orbital parameters, their uncertainty, and the IP value for December 22, 2032. We used observations of this asteroid in the time interval from December 25, 2024 until several epochs in January and February 2025 (see first column in Table~\ref{tab:2024_YR4}). At that time, impact monitoring centers claimed that the impact probability was of the order of several per cent. We grouped all observations as a function of magnitude (or geocentric distance). The division into three magnitude groups was performed automatically using the \texttt{histogram} function in Python, which partitions the full magnitude range into three intervals and provides the corresponding boundaries. Each observation was then assigned to a group according to its magnitude. The number of groups was chosen to ensure at least 30 observations per group for statistical significance, and the grouping procedure was repeated for each run to adapt to the updated dataset. Eventually, there are 3 groups over magnitude range $\approx 16$ to 25, with at least 30 observations per group to ensure statistical significance. Fig.~\ref{fig:groups_of_observations} shows the division of the groups for each dataset.

\begin{figure*}[h!]
\centering
\includegraphics[height=0.95\textheight]{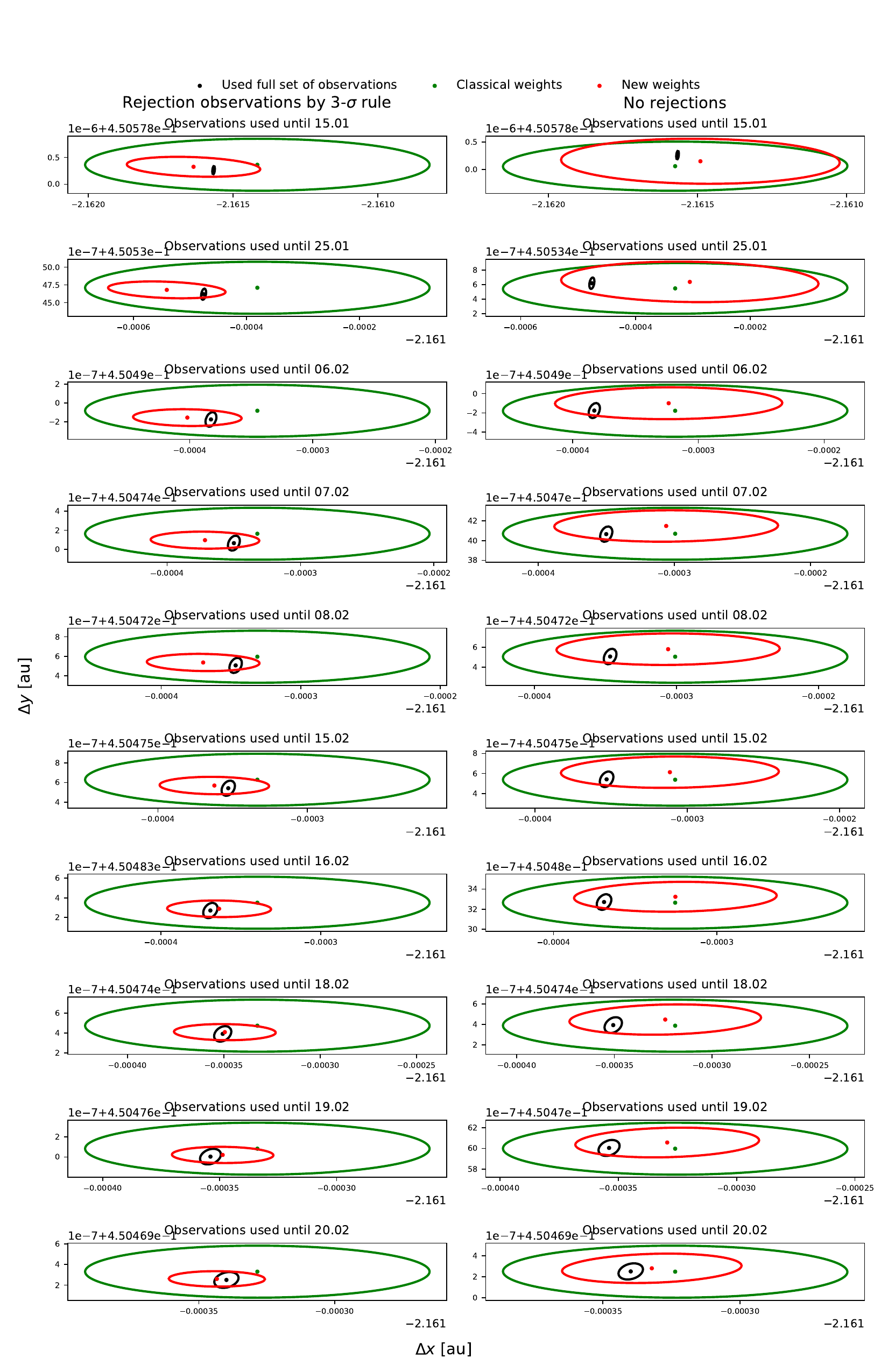}

\caption{Predictions of the position of 2024~YR$_4$ on date of JWST observation, with the $3\sigma$ uncertainty region by different orbits. The left column presents predictions by different orbits with rejection of observations by the $3\sigma$ rule. The right column presents the results without any observation rejection.}
\label{fig:predictions}
\end{figure*}

\begin{table*}[h]
\centering
\caption{Application on asteroid 2024~YR$_4$. }
\label{tab:2024_YR4}
\begin{tabular}{|c|c|c|c|c|c|c|c|c|c|c|}
\hline
 \multirow{2}{*}{until} & \multicolumn{3}{c|}{classical weights} & \multicolumn{6}{c|}{New weights} & \multirow{2}{*}{$\delta \chi^2$}\\ \cline{2-10}
  & $N_{obs_1}$ & $\chi_{cl}^2$ & probability & $N_{obs_2}$ & $\chi_{new}^2$ & probability & $K_1$  & $K_2$ & $K_3$ &  \\ \hline
Jan 15 &	216		& 1.338 & 3.21$\cdot 10^{-3}$ & 	160    & 0.484 & 4.22 $\cdot 10^{-3}$ & 0.247  & 0.253 & 0.170 & 2.762 \\
Jan 25 &	275		& 0.734 & 6.19$\cdot 10^{-3}$ & 	189    & 0.474 & 4.31 $\cdot 10^{-4}$ & 0.264  & 0.158 & 0.173 & 1.549 \\
Feb 06 &	387		& 0.309 & 1.62$\cdot 10^{-2}$ & 	297    & 0.204 & 1.33 $\cdot 10^{-4}$ & 0.271  & 0.149 & 0.247 & 1.518 \\
Feb 07 &	397		& 0.183 & 1.68$\cdot 10^{-2}$ & 	305	   & 0.228 & 2.58 $\cdot 10^{-5}$ & 0.273  & 0.149 & 0.254 & 0.805 \\
Feb 08 &	401		& 0.164 & 1.71$\cdot 10^{-2}$ & 	306	   & 0.230 & 1.44 $\cdot 10^{-5}$ & 0.273  & 0.149 & 0.255 & 0.715 \\
Feb 15 &	402		& 0.182 & 1.89$\cdot 10^{-2}$ & 	308	   & 0.158 & 2.16 $\cdot 10^{-4}$ & 0.285  & 0.149 & 0.259 & 1.156 \\
Feb 16 &	403		& 0.250 & 2.12$\cdot 10^{-2}$ & 	304	   & 0.148 & 3.48 $\cdot 10^{-3}$ & 0.272  & 0.151 & 0.246 & 1.691 \\
Feb 18 &	410		& 0.186 & 2.29$\cdot 10^{-2}$ & 	314	   & 0.149 & 1.88 $\cdot 10^{-4}$ & 0.278  & 0.145 & 0.272 & 1.245 \\
Feb 19 &	421		& 0.207 & 2.76$\cdot 10^{-2}$ & 	324	   & 0.156 & 1.13 $\cdot 10^{-4}$ & 0.289  & 0.146 & 0.272 & 1.334 \\
Feb 20 &	437		& 0.199 & 2.24$\cdot 10^{-2}$ & 	336	   & 0.185 & 2.15 $\cdot 10^{-9}$ & 0.271  & 0.140 & 0.266 & 1.075 \\
\hline
\end{tabular}

\medskip
Comparison of orbits before and after applying the proposed weighting scheme for
different sets of observations (from 25.12.2024 until the epoch given in column 1), when applyin rejection of observations by the 3-$\sigma$ rule. $N_{obs_1}$ and $N_{obs_2}$ are the number of observations used for the orbital fit with classical weights and new weight, respectively.
$\chi_{cl}^2$ and $\chi_{new}^2$ are $\chi^2$ for future
observations for orbits with classical weights and new weights, correspondingly. Probability is the impact probability estimated by semi-linear Partial Banana Mapping method \citep{2025A&A...699A.158V}. Values of $\delta \chi^2 = \chi_{cl}^2 / \chi_{new}^2 $ greater than 1 indicate that the new orbit is in better agreement with future observations. $K_i$ are the $K$ factors for i-th observation group obtained from Eq.~(\ref{eq:kgaia}).

\end{table*}

\begin{table*}[h]
\centering
\caption{Same as Table \ref{tab:2024_YR4}, but without rejection of observations by the 3-$\sigma$ rule.}
\label{tab:2024_YR4_norej}
\begin{tabular}{|c|c|c|c|c|c|c|c|c|c|}
\hline
 \multirow{2}{*}{until} & \multirow{2}{*}{N$_{obs}$} & \multicolumn{2}{c|}{classical weights} & \multicolumn{5}{c|}{New weights} & \multirow{2}{*}{$\delta \chi^2$} \\ \cline{3-9}
 &   & $\chi_{cl}^2$ & probability & $\chi_{new}^2$ & probability & $K_1$  & $K_2$ & $K_3$ &  \\ \hline
Jan 15 &  225	 	& 0.284 & 3.94$\cdot 10^{-3}$ & 0.551 & 4.81 $\cdot 10^{-3}$ & 0.73   & 1.22  &  0.84 & 0.516 \\
Jan 25 &  284	 	& 1.593 & 4.13$\cdot 10^{-3}$ & 2.054 & 1.88 $\cdot 10^{-3}$ & 0.68   & 1.21  & 0.69  & 0.775 \\
Feb 06 &  396	 	& 0.689 & 1.33$\cdot 10^{-2}$ & 0.564 & 1.73 $\cdot 10^{-2}$ & 0.70   & 1.19  & 0.53  & 1.221 \\
Feb 07 &  405	 	& 0.508 & 1.68$\cdot 10^{-2}$ & 0.389 & 2.56 $\cdot 10^{-2}$ & 0.70   & 1.19  & 0.52  & 1.306 \\
Feb 08 &  409	 	& 0.444 & 1.83$\cdot 10^{-2}$ & 0.351 & 2.80 $\cdot 10^{-2}$ & 0.70   & 1.19  & 0.53  & 1.265 \\
Feb 15 &  410	 	& 0.439 & 1.83$\cdot 10^{-2}$ & 0.364 & 2.30 $\cdot 10^{-2}$ & 0.70   & 1.19  & 0.53  & 1.207 \\
Feb 16 &  412	 	& 0.428 & 2.08$\cdot 10^{-2}$ & 0.401 & 3.21 $\cdot 10^{-2}$ & 0.70   & 1.16  & 0.48  & 1.067 \\
Feb 18 &  418	 	& 0.306 & 2.58$\cdot 10^{-2}$ & 0.244 & 4.51 $\cdot 10^{-2}$ & 0.73   & 1.07  & 0.47  & 1.256 \\
Feb 19 &  429	 	& 0.290 & 3.17$\cdot 10^{-2}$ & 0.244 & 5.10 $\cdot 10^{-2}$ & 0.73   & 1.07  & 0.46  & 1.189 \\
Feb 20 &  445	 	& 0.236 & 2.79$\cdot 10^{-2}$ & 0.192 & 6.17 $\cdot 10^{-3}$ & 0.82   & 0.99  & 0.45  & 1.230 \\
\hline
\end{tabular}
\end{table*}



During the orbital fitting, we rejected observations whose residuals exceeded three times their estimated uncertainty (the $3\sigma$ rule). Therefore, the $K$-factors from Eq.~\ref{eq:k_factors} were also estimated using the reduced set of observations and must be corrected accordingly. Since $K_i$ represents an unbiased estimate of the standard deviation of the $i$-th group, we divide this value by 0.9866 to recover the corresponding non-truncated standard deviation \citep{johnson1994continuous}. This correction changed the values of $K$ by approximately $1.3\%$, but it had no noticeable effect on the resulting orbital solutions or on the subsequent analysis.

The comparison of the results is presented in Table~\ref{tab:2024_YR4}.  In seven out of ten cases (Jan 15, Jan 25, Feb 6, Feb 15, Feb 16, Feb 18, and Feb 19), the new weighting scheme produced more precise orbital solutions. In two cases (Feb 7 and Feb 8), the orbits computed with classical weights better reproduced future observations, while for Feb 20 -- with the largest observational arc -- the difference is negligible. Interestingly, the derived $K$-factors are significantly lower than unity, indicating that the observations are underweighted (i.e., their uncertainties are overestimated). As a result, more measurements are rejected under the $3\sigma$ criterion when using the new scheme. 

We have also computed the impact probabilities for each orbit using the semi-linear Impactor Partial Banana Mapping method~\citep{2025A&A...699A.158V}. The method uses the curvilinear coordinate system \citep{2015MNRAS.446..705V} to accurately represent the uncertainty region of an asteroid, finds the closest virtual asteroid on the main line of the uncertainty region \citep{2020MNRAS.492.4546V} and then explicitly finds the impacting orbital solution.
After applying the new weighting scheme, the formal errors of the orbital parameters decrease substantially, which likely led to lower impact probability estimates. Indeed, with the new orbits the impact probabilities do not exceed 0.5\%, remaining well below the 1\% IAWN  alert threshold.

Table~\ref{tab:2024_YR4_norej} presents the same comparison but without applying any observation rejection during the orbital fit. The results  are similar, the new weighting solutions are in general more precise in forward propagation. The $K$ values are now closer to 1, compared to the one with outlier rejection. This is because the low precision observations are not rejected and consequently increase $K$-factors in Eq.~(\ref{eq:k_factors}). The uncertainty regions are still smaller with new weights, but not significantly, which yields minor difference in the IP values between the new-weighting and the classical-weighting orbits. The $\chi^2_{new}$ values are larger than for the case where the rejection by 3$\sigma$ rule is applied, meaning the outlier rejection led to more accurate orbital parameters. 

Figure~\ref{fig:predictions} illustrates an alternative comparison of orbital precision. For each orbit, we propagated the state vector to May 11, 2025, the epoch of the last available observation of 2024~YR$_4$ obtained by the James Webb Space Telescope (JWST). Thanks to the high precision from JWST, the position at this epoch is well constrained, and it serves as a good reference. The figure shows that this position always lies within the $3\sigma$ uncertainty ellipsoid of each considered orbit. Nevertheless, the uncertainty regions derived using the new weights are systematically smaller. Even in cases where the classical-weight predictions lie slightly closer to the true position, the new-weight solutions exhibit smaller uncertainty volumes and, therefore, produce higher precision predictions. It is also worth noting that the application of the $3\sigma$ rejection rule results in substantially more accurate orbits with the new weighting scheme, despite a significant number of observations being discarded.

Although the recent observations ruled out a possible impact of 2024~YR$_4$ with the Earth on December 22, 2032, the asteroid now shows a non-negligible collision probability with the Moon on that date. When using all available observations, the classical weighting scheme yields an impact probability by the Monte-Carlo method of $2.46\%$, with  $0.23\%$ $3\sigma$ confidence interval, while the new weighting scheme gives $(3.65 \pm 0.27)\%$.
Even though a collision with the Earth in 2032 is no longer possible, our analysis reveals a small impact probability for December 22, 2047. Using the classical weighting scheme, two potential impacts were identified with probabilities of $6.07\cdot10^{-5}$ and $4.42\cdot10^{-6}$, whereas the new weighting scheme produced a single impact solution with a probability of $1.63\cdot10^{-4}$ (computed by IPBM).

\section{Conclusions}
As the weighted least squares method is widely used in data fitting in the field of orbit determination of asteroids, while the uncertainty of different data points is less well known (or often over/under-estimated), we propose a method of adjusting the weights to better balance and weight the data. The method requires the division of observations by different groups and introduces a scaling factor for the weights within each group. The scaling factors are estimated based on their dispersion, or how well the observations in the group fit the model. 

We do not provide strict recommendations here on how observations should be divided into groups, although some practical constraints apply. Each group must contain a sufficiently large number of measurements; in general, we recommend at least about 30 observations per group to retrieve significant statistics. The key assumption of our method is that the relative weighting within a given group is more reliable, whereas the weight scaling between different groups is uncertain. Therefore, it is natural to define groups based on observational characteristics such as the data type, the observing facility, or similar factors. In this work, we demonstrated two possible grouping strategies for asteroid astrometry. In the first case, we treated \textit{Gaia} measurements as one group and all ground-based observations as another, based on their relative formal astrometric precision. In the second case, we divided the ground-based data simply according to the target’s visual magnitude, again should reflect different formal precision.

To validate the approach, we applied the reweighting scheme to a set of 15 asteroids observed by both ground-based telescopes and by the Gaia mission. The results demonstrate that the new weighting method consistently improves or maintains the quality of orbital fits. In several cases, particularly for asteroids with mixed-quality datasets, the formal errors of the orbital elements decreased significantly. The analysis of asteroid (21) Lutetia provided a striking example, where the proposed method effectively accounted for the biases affecting Gaia observations of bright and extended bodies, yielding a much better fit to the complete observation set. These tests confirm that the scaling factors derived for distinct data subsets can be used to obtain a more robust and reliable orbit determination.

We have further demonstrated the applicability of the new weighting scheme approach in a practical scenario involving the recently discovered near-Earth asteroid 2024~YR$_4$. This object initially attracted attention due to an apparent impact probability of a few percent with the Earth in 2032. Using our new scheme, we computed the orbit and its associated uncertainties for several observational intervals. In most cases, the resulting orbits were more precise, yielding to a smaller uncertainty regions than those derived with classical weights. Consequently, the estimated impact probabilities were systematically reduced and remained below the 1\% IAWN alert threshold. Even in cases where the predictions based on classical weights appeared closer to subsequent observations, the new-weight orbits provided more reliable prediction estimates.


\section*{Acknowledgements}
This project has received funding from the European Union’s Horizon 2020 research and innovation programme under the Marie Sklodowska-Curie grant agreement \#101068341 “NEOForCE”.
This work has received support from France 2030 through the project named Académie Spatiale d'Île-de-France (https://academiespatiale.fr/) managed by the National Research Agency under bearing the reference ANR-23-CMAS-0041.
This material is based upon work supported by the National Science Foundation under Grant No. (2307569). This research award is partially funded by a generous gift of Charles Simonyi to the NSF Division of Astronomical Sciences. The award is made in recognition of significant contributions to Rubin Observatory’s Legacy Survey of Space and Time. Any opinions, findings, and conclusions or recommendations expressed in this material are those of the author(s) and do not necessarily reflect the views of the National Science Foundation.

\section*{Contribution}

D. E. Vavilov: Conceptualization, Methodology, Investigation, Validation, Formal analysis, Software (impact probability code, implementation), Supervision, Visualization, Writing Original draft, Review \& Editing.
Z. Liu: Methodology, Investigation, Validation, Formal analysis, Software (implementation), Visualization, Writing Original draft, Review \& Editing.
D. Hestroffer: Methodology, Supervision, Writing, Review \& Editing.
J. Desmars: Software (orbit determination code).



\bibliographystyle{icarus} 

\begin{thebibliography}{}

\bibitem[{Carpino} et~al.(2003){Carpino}, {Milani}, and
  {Chesley}]{2003Icar..166..248C}
{Carpino}, M., {Milani}, A., {Chesley}, S.~R., 2003.
\newblock {Error statistics of asteroid optical astrometric observations}.
\newblock \icarus~166 (2), 248--270.

\bibitem[{Desmars}(2015){Desmars}]{NIMA}
{Desmars}, J., 2015.
\newblock {Detection of Yarkovsky acceleration in the context of precovery
  observations and the future Gaia catalogue}.
\newblock \aap~575, A53.

\bibitem[{Farnocchia} et~al.(2015){Farnocchia}, {Chesley}, {Chamberlin}, and
  {Tholen}]{2015Icar..245...94F}
{Farnocchia}, D., {Chesley}, S.~R., {Chamberlin}, A.~B., {Tholen}, D.~J., 2015.
\newblock {Star catalog position and proper motion corrections in asteroid
  astrometry}.
\newblock \icarus~245, 94--111.

\bibitem[{Gaia Collaboration} et~al.(2023){Gaia Collaboration}, {David},
  {Mignard}, {Hestroffer}, {Tanga}, {Spoto}, {Berthier}, {Pauwels}, {Roux},
  {Barbier}, {Cellino}, {Carry}, {Delbo}, {Dell'Oro}, {Fouron}, {Galluccio},
  {Klioner}, {Mary}, {Muinonen}, {Ordenovic}, {Oreshina-Slezak}, {Panem},
  {Petit}, {Portell}, {Brown}, {Thuillot}, {Vallenari}, {Prusti}, {de Bruijne},
  {Arenou}, {Babusiaux}, {Biermann}, {Creevey}, {Ducourant}, {Evans}, {Eyer},
  {Guerra}, {Hutton}, {Jordi}, {Lammers}, {Lindegren}, {Luri}, {Randich},
  {Sartoretti}, {Smiljanic}, {Walton}, {Bailer-Jones}, {Bastian}, {Cropper},
  {Drimmel}, {Katz}, {Soubiran}, {van Leeuwen}, {Audard}, {Bakker}, {Blomme},
  {Casta{\~n}eda}, {De Angeli}, {Fabricius}, {Fouesneau}, {Fr{\'e}mat},
  {Guerrier}, {Masana}, {Messineo}, {Nicolas}, {Nienartowicz}, {Pailler},
  {Panuzzo}, {Riclet}, {Seabroke}, {Sordo}, {Th{\'e}venin}, {Gracia-Abril},
  {Teyssier}, {Altmann}, {Benson}, {Burgess}, {Busonero}, {Busso},
  {C{\'a}novas}, {Cheek}, {Clementini}, {Damerdji}, {Davidson}, {de Teodoro},
  {Delchambre}, {Fraile Garcia}, {Garabato}, {Garc{\'\i}a-Lario}, {Garralda
  Torres}, {Gavras}, {Haigron}, {Hambly}, {Harrison}, {Hatzidimitriou},
  {Hern{\'a}ndez}, {Hodgkin}, {Holl}, {Jamal}, {Jordan}, {Krone-Martins},
  {Lanzafame}, {L{\"o}ffler}, {Lorca}, {Marchal}, {Marrese}, {Moitinho},
  {Nu{\~n}ez Campos}, {Osborne}, {Pancino}, {Recio-Blanco}, {Riello},
  {Rimoldini}, {Robin}, {Roegiers}, {Sarro}, {Schultheis}, {Siopis}, {Smith},
  {Sozzetti}, {Utrilla}, {van Leeuwen}, {Weingrill}, {Abbas},
  {{\'A}brah{\'a}m}, {Abreu Aramburu}, {Aerts}, {Altavilla}, {{\'A}lvarez},
  {Alves}, {Anderson}, {Antoja}, {Baines}, {Baker}, {Balog}, {Barache},
  {Barbato}, {Barros}, {Barstow}, {Bartolom{\'e}}, {Bashi}, {Bauchet},
  {Baudeau}, {Becciani}, {Bedin}, {Bellas-Velidis}, {Bellazzini}, {Beordo},
  {Berihuete}, {Bernet}, {Bertolotto}, {Bertone}, {Bianchi}, {Binnenfeld},
  {Blazere}, {Boch}, {Bombrun}, {Bouquillon}, {Bragaglia}, {Braine},
  {Bramante}, {Breedt}, {Bressan}, {Brouillet}, {Brugaletta}, {Bucciarelli},
  {Butkevich}, {Buzzi}, {Caffau}, {Cancelliere}, {Cannizzo}, {Carballo},
  {Carlucci}, {Carnerero}, {Carrasco}, {Carretero}, {Carton}, {Casamiquela},
  {Castellani}, {Castro-Ginard}, {Cesare}, {Charlot}, {Chemin}, {Chiaramida},
  {Chiavassa}, {Chornay}, {Collins}, {Contursi}, {Cooper}, {Cornez}, {Crosta},
  {Crowley}, {Dafonte}, {de Laverny}, {De Luise}, {De March}, {de Souza}, {de
  Torres}, {del Peloso}, {Delgado}, {Dharmawardena}, {Diakite}, {Diener},
  {Distefano}, {Dolding}, {Dsilva}, {Dur{\'a}n}, {Enke}, {Esquej}, {Fabre},
  {Fabrizio}, {Faigler}, {Fatovi{\'c}}, {Fedorets},
  {Fern{\'a}ndez-Hern{\'a}ndez}, {Fernique}, {Figueras}, {Fournier}, {Gai},
  {Galinier}, {Garcia-Gutierrez}, {Garc{\'\i}a-Torres}, {Garofalo}, {Gerlach},
  {Geyer}, {Giacobbe}, {Gilmore}, {Girona}, {Giuffrida}, {Gomel}, {Gomez},
  {Gonz{\'a}lez-N{\'u}{\~n}ez}, {Gonz{\'a}lez-Santamar{\'\i}a}, {Gosset},
  {Granvik}, {Gregori Barrera}, {Guti{\'e}rrez-S{\'a}nchez}, {Haywood},
  {Helmer}, {Helmi}, {Henares}, {Hidalgo}, {Hilger}, {Hobbs}, {Hottier},
  {Huckle}, {Jab{\l}o{\'n}ska}, {Jansen}, {Jim{\'e}nez-Arranz}, {Juaristi
  Campillo}, {Khanna}, {Kordopatis}, {K{\'o}sp{\'a}l}, {Kostrzewa-Rutkowska},
  {Kun}, {Lambert}, {Lanza}, {Le Campion}, {Lebreton}, {Lebzelter}, {Leccia},
  {Lecoeur-Taibi}, {Lecoutre}, {Liao}, {Liberato}, {Licata}, {Lindstr{\o}m},
  {Lister}, {Livanou}, {Lobel}, {Loup}, {Mahy}, {Mann}, {Manteiga}, {Marchant},
  {Marconi}, {Mar{\'\i}n Pina}, {Marinoni}, {Marshall}, {Mart{\'\i}n Lozano},
  {Mart{\'\i}n-Fleitas}, {Marton}, {Masip}, {Massari}, {Mastrobuono-Battisti},
  {Mazeh}, {McMillan}, {Meichsner}, {Messina}, {Michalik}, {Millar}, {Mints},
  {Molina}, {Molinaro}, {Moln{\'a}r}, {Monari}, {Mongui{\'o}}, {Montegriffo},
  {Montero}, {Mor}, {Mora}, {Morbidelli}, {Morel}, {Morris}, {Mowlavi},
  {Munoz}, {Muraveva}, {Murphy}, {Musella}, {Nagy}, {Nieto}, {Noval}, {Ogden},
  {Pagani}, {Pagano}, {Palaversa}, {Palicio}, {Pallas-Quintela}, {Panahi},
  {Payne-Wardenaar}, {Pegoraro}, {Penttil{\"a}}, {Pesciullesi}, {Piersimoni},
  {Pinamonti}, {Pineau}, {Plachy}, {Plum}, {Poggio}, {Pourbaix}, {Pr{\v{s}}a},
  {Pulone}, {Racero}, {Rainer}, {Raiteri}, {Ramos}, {Ramos-Lerate},
  {Ratajczak}, {Re Fiorentin}, {Regibo}, {Reyl{\'e}}, {Ripepi}, {Riva}, {Rix},
  {Rixon}, {Robichon}, {Robin}, {Romero-G{\'o}mez}, {Rowell}, {Royer}, {Ruz
  Mieres}, {Rybicki}, {Sadowski}, {S{\'a}ez N{\'u}{\~n}ez}, {Sagrist{\`a}
  Sell{\'e}s}, {Sahlmann}, {Sanchez Gimenez}, {Sanna}, {Santove{\~n}a},
  {Sarasso}, {Sarrate Riera}, {Sciacca}, {Segovia}, {S{\'e}gransan}, {Shahaf},
  {Siebert}, {Siltala}, {Slezak}, {Smart}, {Snaith}, {Solano}, {Solitro},
  {Souami}, {Souchay}, {Spina}, {Spitoni}, {Squillante}, {Steele},
  {Steidelm{\"u}ller}, {Surdej}, {Szabados}, {Taris}, {Taylor}, {Teixeira},
  {Tisani{\'c}}, {Tolomei}, {Torra}, {Torralba Elipe}, {Trabucchi}, {Tsantaki},
  {Ulla}, {Unger}, {Vanel}, {Vecchiato}, {Vicente}, {Voutsinas}, {Weiler},
  {Wyrzykowski}, {Zhao}, {Zorec}, {Zwitter}, {Balaguer-N{\'u}{\~n}ez},
  {Leclerc}, {Morgenthaler}, {Robert}, and {Zucker}]{david23}
{Gaia Collaboration}, et~al., 2023.
\newblock {Gaia Focused Product Release: Asteroid orbital solution. Properties
  and assessment}.
\newblock \aap~680, A37.

\bibitem[{Gaia Collaboration} et~al.(2018){Gaia Collaboration}, {Spoto},
  {Tanga}, {Mignard}, {Berthier}, {Carry}, {Cellino}, {Dell'Oro}, {Hestroffer},
  {Muinonen}, {Pauwels}, {Petit}, {David}, {De Angeli}, {Delbo},
  {Fr{\'e}zouls}, {Galluccio}, {Granvik}, {Guiraud}, {Hern{\'a}ndez},
  {Ord{\'e}novic}, {Portell}, {Poujoulet}, {Thuillot}, {Walmsley}, {Brown},
  {Vallenari}, {Prusti}, {de Bruijne}, {Babusiaux}, {Bailer-Jones}, {Biermann},
  {Evans}, {Eyer}, {Jansen}, {Jordi}, {Klioner}, {Lammers}, {Lindegren},
  {Luri}, {Panem}, {Pourbaix}, {Randich}, {Sartoretti}, {Siddiqui}, {Soubiran},
  {van Leeuwen}, {Walton}, {Arenou}, {Bastian}, {Cropper}, {Drimmel}, {Katz},
  {Lattanzi}, {Bakker}, {Cacciari}, {Casta{\~n}eda}, {Chaoul}, {Cheek},
  {Fabricius}, {Guerra}, {Holl}, {Masana}, {Messineo}, {Mowlavi},
  {Nienartowicz}, {Panuzzo}, {Riello}, {Seabroke}, {Th{\'e}venin},
  {Gracia-Abril}, {Comoretto}, {Garcia-Reinaldos}, {Teyssier}, {Altmann},
  {Andrae}, {Audard}, {Bellas-Velidis}, {Benson}, {Blomme}, {Burgess}, {Busso},
  {Clementini}, {Clotet}, {Creevey}, {Davidson}, {De Ridder}, {Delchambre},
  {Ducourant}, {Fern{\'a}ndez-Hern{\'a}ndez}, {Fouesneau}, {Fr{\'e}mat},
  {Garc{\'\i}a-Torres}, {Gonz{\'a}lez-N{\'u}{\~n}ez}, {Gonz{\'a}lez-Vidal},
  {Gosset}, {Guy}, {Halbwachs}, {Hambly}, {Harrison}, {Hodgkin}, {Hutton},
  {Jasniewicz}, {Jean-Antoine-Piccolo}, {Jordan}, {Korn}, {Krone-Martins},
  {Lanzafame}, {Lebzelter}, {L{\"o}}, {Manteiga}, {Marrese},
  {Mart{\'\i}n-Fleitas}, {Moitinho}, {Mora}, {Osinde}, {Pancino},
  {Recio-Blanco}, {Richards}, {Rimoldini}, {Robin}, {Sarro}, {Siopis}, {Smith},
  {Sozzetti}, {S{\"u}veges}, {Torra}, {van Reeven}, {Abbas}, {Abreu Aramburu},
  {Accart}, {Aerts}, {Altavilla}, {{\'A}lvarez}, {Alvarez}, {Alves},
  {Anderson}, {Andrei}, {Anglada Varela}, {Antiche}, {Antoja}, {Arcay},
  {Astraatmadja}, {Bach}, {Baker}, {Balaguer-N{\'u}{\~n}ez}, {Balm}, {Barache},
  {Barata}, {Barbato}, {Barblan}, {Barklem}, {Barrado}, {Barros}, {Barstow},
  {Bartholom{\'e} Mu{\~n}oz}, {Bassilana}, {Becciani}, {Bellazzini},
  {Berihuete}, {Bertone}, {Bianchi}, {Bienaym{\'e}}, {Blanco-Cuaresma}, {Boch},
  {Boeche}, {Bombrun}, {Borrachero}, {Bossini}, {Bouquillon}, {Bourda},
  {Bragaglia}, {Bramante}, {Breddels}, {Bressan}, {Brouillet},
  {Br{\"u}semeister}, {Brugaletta}, {Bucciarelli}, {Burlacu}, {Busonero},
  {Butkevich}, {Buzzi}, {Caffau}, {Cancelliere}, {Cannizzaro}, {Cantat-Gaudin},
  {Carballo}, {Carlucci}, {Carrasco}, {Casamiquela}, {Castellani},
  {Castro-Ginard}, {Charlot}, {Chemin}, {Chiavassa}, {Cocozza}, {Costigan},
  {Cowell}, {Crifo}, {Crosta}, {Crowley}, {Cuypers}, {Dafonte}, {Damerdji},
  {Dapergolas}, {David}, {de Laverny}, {De Luise}, {De March}, {de Souza}, {de
  Torres}, {Debosscher}, {del Pozo}, {Delgado}, {Delgado}, {Diakite}, {Diener},
  {Distefano}, {Dolding}, {Drazinos}, {Dur{\'a}n}, {Edvardsson}, {Enke},
  {Eriksson}, {Esquej}, {Eynard Bontemps}, {Fabre}, {Fabrizio}, {Faigler},
  {Falc{\~a}o}, {Farr{\`a}s Casas}, {Federici}, {Fedorets}, {Fernique},
  {Figueras}, {Filippi}, {Findeisen}, {Fonti}, {Fraile}, {Fraser}, {Gai},
  {Galleti}, {Garabato}, {Garc{\'\i}a-Sedano}, {Garofalo}, {Garralda}, {Gavel},
  {Gavras}, {Gerssen}, {Geyer}, {Giacobbe}, {Gilmore}, {Girona}, {Giuffrida},
  {Glass}, {Gomes}, {Gueguen}, {Guerrier}, {Guti{\'e}}, {Haigron},
  {Hatzidimitriou}, {Hauser}, {Haywood}, {Heiter}, {Helmi}, {Heu}, {Hilger},
  {Hobbs}, {Hofmann}, {Holland}, {Huckle}, {Hypki}, {Icardi}, {Jan{\ss}en},
  {Jevardat de Fombelle}, {Jonker}, {Juh{\'a}sz}, {Julbe}, {Karampelas},
  {Kewley}, {Klar}, {Kochoska}, {Kohley}, {Kolenberg}, {Kontizas}, {Kontizas},
  {Koposov}, {Kordopatis}, {Kostrzewa-Rutkowska}, {Koubsky}, {Lambert},
  {Lanza}, {Lasne}, {Lavigne}, {Le Fustec}, {Le Poncin-Lafitte}, {Lebreton},
  {Leccia}, {Leclerc}, {Lecoeur-Taibi}, {Lenhardt}, {Leroux}, {Liao}, {Licata},
  {Lindstr{\o}m}, {Lister}, {Livanou}, {Lobel}, {L{\'o}pez}, {Managau}, {Mann},
  {Mantelet}, {Marchal}, {Marchant}, {Marconi}, {Marinoni}, {Marschalk{\'o}},
  {Marshall}, {Martino}, {Marton}, {Mary}, {Massari}, {Matijevi{\v{c}}},
  {Mazeh}, {McMillan}, {Messina}, {Michalik}, {Millar}, {Molina}, {Molinaro},
  {Moln{\'a}r}, {Montegriffo}, {Mor}, {Morbidelli}, {Morel}, {Morris},
  {Mulone}, {Muraveva}, {Musella}, {Nelemans}, {Nicastro}, {Noval},
  {O'Mullane}, {Ord{\'o}{\~n}ez-Blanco}, {Osborne}, {Pagani}, {Pagano},
  {Pailler}, {Palacin}, {Palaversa}, {Panahi}, {Pawlak}, {Piersimoni},
  {Pineau}, {Plachy}, {Plum}, {Poggio}, {Pr{\v{s}}a}, {Pulone}, {Racero},
  {Ragaini}, {Rambaux}, {Ramos-Lerate}, {Regibo}, {Reyl{\'e}}, {Riclet},
  {Ripepi}, {Riva}, {Rivard}, {Rixon}, {Roegiers}, {Roelens},
  {Romero-G{\'o}mez}, {Rowell}, {Royer}, {Ruiz-Dern}, {Sadowski}, {Sagrist{\`a}
  Sell{\'e}s}, {Sahlmann}, {Salgado}, {Salguero}, {Sanna}, {Santana-Ros},
  {Sarasso}, {Savietto}, {Schultheis}, {Sciacca}, {Segol}, {Segovia},
  {S{\'e}gransan}, {Shih}, {Siltala}, {Silva}, {Smart}, {Smith}, {Solano},
  {Solitro}, {Sordo}, {Soria Nieto}, {Souchay}, {Spagna}, {Stampa}, {Steele},
  {Steidelm{\"u}ller}, {Stephenson}, {Stoev}, {Suess}, {Surdej}, {Szabados},
  {Szegedi-Elek}, {Tapiador}, {Taris}, {Tauran}, {Taylor}, {Teixeira},
  {Terrett}, {Teyssandier}, {Titarenko}, {Torra Clotet}, {Turon}, {Ulla},
  {Utrilla}, {Uzzi}, {Vaillant}, {Valentini}, {Valette}, {van Elteren}, {Van
  Hemelryck}, {van Leeuwen}, {Vaschetto}, {Vecchiato}, {Veljanoski}, {Viala},
  {Vicente}, {Vogt}, {von Essen}, {Voss}, {Votruba}, {Voutsinas}, {Weiler},
  {Wertz}, {Wevers}, {Wyrzykowski}, {Yoldas}, {{\v{Z}}erjal}, {Ziaeepour},
  {Zorec}, {Zschocke}, {Zucker}, {Zurbach}, and {Zwitter}]{gaiadr2}
{Gaia Collaboration}, et~al., 2018.
\newblock {Gaia Data Release 2. Observations of solar system objects}.
\newblock \aap~616, A13.

\bibitem[{Gauss}(1809){Gauss}]{1809tmcc.book.....G}
{Gauss}, K.~F., 1809.
\newblock {\em {Theoria motvs corporvm coelestivm in sectionibvs conicis solem
  ambientivm.}}

\bibitem[Johnson et~al.(1994)Johnson, Kotz, and
  Balakrishnan]{johnson1994continuous}
Johnson, N.~L., Kotz, S., Balakrishnan, N., 1994.
\newblock {\em Continuous Univariate Distributions, Volume 1}.
\newblock Wiley.

\bibitem[{Muinonen} et~al.(2022){Muinonen}, {Berthier}, {Cellino}, {David}, {De
  Angeli}, {Delb{\'o}}, {Dell-Oro}, {Galluccio}, {Hestroffer}, {Mignard},
  {Pauwels}, {Spoto}, and {Tanga}]{gdr3-22}
{Muinonen}, K., et~al., 2022.
\newblock {Gaia DR3 documentation Chapter 8: Solar System Objects}.
\newblock Gaia DR3 documentation, European Space Agency; Gaia Data Processing
  and Analysis Consortium. Online at
  https://gea.esac.esa.int/archive/documentation/GDR3/index.html, id. 8.

\bibitem[{Tanga} et~al.(2023){Tanga}, {Pauwels}, {Mignard}, {Muinonen},
  {Cellino}, {David}, {Hestroffer}, {Spoto}, {Berthier}, {Guiraud}, {Roux},
  {Carry}, {Delbo}, {Dell'Oro}, {Fouron}, {Galluccio}, {Jonckheere}, {Klioner},
  {Lefustec}, {Liberato}, {Ord{\'e}novic}, {Oreshina-Slezak}, {Penttil{\"a}},
  {Pailler}, {Panem}, {Petit}, {Portell}, {Poujoulet}, {Thuillot}, {Van
  Hemelryck}, {Burlacu}, {Lasne}, and {Managau}]{tanga23}
{Tanga}, P., et~al., 2023.
\newblock {Gaia Data Release 3. The Solar System survey}.
\newblock \aap~674, A12.

\bibitem[{Vavilov}(2020){Vavilov}]{2020MNRAS.492.4546V}
{Vavilov}, D.~E., 2020.
\newblock {The partial banana mapping: a robust linear method for impact
  probability estimation}.
\newblock \mnras~492 (3), 4546--4552.

\bibitem[{Vavilov} and {Hestroffer}(2024){Vavilov} and
  {Hestroffer}]{2024A&A...689A..49V}
{Vavilov}, D.~E., {Hestroffer}, D., 2024.
\newblock {Asteroid follow-up and precovery problem: Partial banana mapping
  solution}.
\newblock \aap~689, A49.

\bibitem[{Vavilov} and {Hestroffer}(2025){Vavilov} and
  {Hestroffer}]{2025A&A...699A.158V}
{Vavilov}, D.~E., {Hestroffer}, D., 2025.
\newblock {Semilinear impact monitoring: Partial-banana mapping: Search for
  impactors}.
\newblock A\&A~699, A158.

\bibitem[{Vavilov} and {Medvedev}(2015){Vavilov} and
  {Medvedev}]{2015MNRAS.446..705V}
{Vavilov}, D.~E., {Medvedev}, Y.~D., 2015.
\newblock {A fast method for estimation of the impact probability of near-Earth
  objects}.
\newblock \mnras~446 (1), 705--709.

\bibitem[{Vere{\v{s}}} et~al.(2017){Vere{\v{s}}}, {Farnocchia}, {Chesley}, and
  {Chamberlin}]{2017Icar..296..139V}
{Vere{\v{s}}}, P., {Farnocchia}, D., {Chesley}, S.~R., {Chamberlin}, A.~B.,
  2017.
\newblock {Statistical analysis of astrometric errors for the most productive
  asteroid surveys}.
\newblock \icarus~296, 139--149.

\end{thebibliography}







\end{document}